\documentclass[12pt]{article}
\usepackage[english]{babel}
\usepackage{amsmath}
\usepackage{cite}
\begin{document}

\title{\bf{}BRST approach to Lagrangian formulation of bosonic totally antisymmeric tensor fields in curved
space}

\author{\sc I.L. Buchbinder${}^{a}$\thanks{joseph@tspu.edu.ru},
V.A. Krykhtin${}^{a,b}$\thanks{krykhtin@tspu.edu.ru}, L.L.
Ryskina$^a$\thanks{ryskina@tspu.edu.ru}
\\[0.5cm]
\it ${}^a$Department of Theoretical Physics,\\
\it Tomsk State Pedagogical University,\\
\it Tomsk 634061, Russia\\[0.3cm]
\it ${}^b$Laboratory of Mathematical Physics,\\
\it Tomsk Polytechnic University,\\
\it Tomsk 634050, Russia}

\date{}

\maketitle
\thispagestyle{empty}

\begin{abstract}
We apply the BRST approach, previously developed for higher spin
field theories, to gauge invariant Lagrangian construction for
antisymmetric massive and massless bosonic fields in arbitrary
d-dimensional curved space. The obtained theories are reducible
gauge models both in massless and massive cases and the order of
reducibility grows with the value of the rank of the antisymmetric
field. In both the cases the Lagrangians contain the sets of
auxiliary fields and possess more rich gauge symmetry in comparison
with standard Lagrangian formulation for the antisymmetric fields.
This serves additional demonstration of universality of the BRST
approach for Lagrangian constructions in various field models.
\end{abstract}

\section{Introduction}

BRST-BFV- construction \cite{BRST}, which was initially developed
for quantization of gauge theories, turned out to be the power
method for deriving the Lagrange formulation in higher spin field
theory\footnote{Aspects of modern state of higher spin field theory
are discussed in the reviews \cite{reviews}.}. Indeed, this method
has successfully been applied to finding the Lagrangians of massless
and massive, bosonic and fermionic higher spin fields with various
symmetry structure of indices in Minkowski and AdS spaces
\cite{massless-bos,boz-ferm,
massless-ferm,massive-bos,massive-ferm,mixed}\footnote{Recently this
approach was applied to Lagrangian formulation of interacting
bosonic higher spin gauge fields \cite{INT}.}. This method, named
BRST-approach to higher spin field theory, in all cases yields to
gauge invariant action in terms of off-shell totally unconstrained
fields and gauge parameters.

The BRST-approach begins with postulating the operator constraints,
determining the irreducible representation of Poincare or AdS
algebra with given spin and constructing the closed algebra of these
operators. It should be pointed out that closing the algebra
requires to introduce some new operators, which can not be
interpreting as the constraints. Also, closing the algebra imposes
the restrictions on space-time geometry and it turns out to be that
in general the scheme under considerations works only for constant
curvature spaces\footnote{Of course in case of spins $s=0,
\frac{1}{2}, 1$ the above algebra is closed for arbitrary space-time
geometry.}. As we will see, there is a non-trivial case when this
scheme successfully works for arbitrary d-dimensional curved
space.

In this paper we develop the BRST-approach to constructing the
Lagrangian formulation for massless and massive totally
antisymmetric bosonic fields in arbitrary curved space\footnote{The
totally antisymmetric fields are the partial case of arbitrary mixed
symmetry higher spin fields. Aspects of Lagrangian formulation for
such fields are discussed in the recent papers
\cite{Brink,Zinoviev,Metsaev,Metsaev-2,Vasiliev,Skvortsov}.}. We show that
the general procedure, described in
\cite{massless-bos,massless-ferm,massive-bos,massive-ferm,mixed}
does not impose any restrictions on
space-time geometry and yields to gauge invariant model containing,
besides the basic field $\varphi_{\mu_1\ldots\mu_p}$, some number of
auxiliary fields. After eliminating the auxiliary fields the
obtained formulation coincides with standard one.

As is well known an antisymmetric bosonic field of rank-$p$
$\varphi_{\mu_1\ldots\mu_p}$
will realize irreducible representation of the Poincare group (in
Minkowski spacetime) if the following equations are
satisfied\footnote{The traceless condition, which is important for
general-type higher spin fields,  be trivial for antisymmetric
fields. We use the metric with mostly plus signature.}
\begin{eqnarray}
(\partial^2-m^2)\varphi_{\mu_1\ldots\mu_p}=0,
&\qquad&
\partial^{\mu_1}\varphi_{\mu_1\ldots\mu_p}=0.
\label{EM-flat}
\end{eqnarray}
When we turn to an arbitrary curved spacetime we suppose that conditions on
$\varphi_{\mu_1\ldots\mu_p}$ which must be satisfied,
tend to (\ref{EM-flat})
in flat space limit.
It tells us that the equations on $\varphi_{\mu_1\ldots\mu_p}$
in curved spacetime must be of the form
\begin{eqnarray}
(\nabla^2-m^2)\varphi_{\mu_1\ldots\mu_p}
+\,\mbox{\it terms with curvature}=0,
&\qquad&
\nabla^{\mu_1}\varphi_{\mu_1\ldots\mu_p}=0,
\label{EM-curved}
\end{eqnarray}
We will see that the ``terms with curvature'' are to be defined
uniquely in process of Lagrangian construction.

The paper is organized as follows. In Section 2 we develop the BRST
approach for massless antisymmetric bosonic fields. In Section 3 we
consider this approach for massive antisymmetric bosonic fields.
Section 4 is devoted to discussion of the results.

\section{Lagrangian construction for massless fields}

To avoid explicit manipulations with a big number of indices it is
convenient to introduce the Fock space generated by fermionic
creation and annihilation operators with tangent space indices
\begin{eqnarray}
\label{comrel}
\{a_a,a_b^+\}=\eta_{ab},
&\qquad&
\eta_{ab}=diag(-,+,+,\cdots,+).
\end{eqnarray}
As usual the tangent space indices and the curved indices are
converted one into another with the help of the vielbein $e^a_\mu$
which is assumed to satisfy the relation
$\nabla_\mu{}e^a_\nu=0$.
Then
one introduces a derivative operator
\begin{eqnarray}
D_\mu=\partial_\mu+\omega_\mu{}^{ab}a_a^+a_b,
&\qquad&
D_\mu|0\rangle=
\partial_\mu|0\rangle=0
\end{eqnarray}
which acts on an arbitrary state vector in this Fock space
\begin{eqnarray}
|\varphi\rangle
&=&
\sum_{p=0}^{}\varphi_{\mu_1\ldots\mu_p}(x)\;
a^{\mu_1+}\ldots a^{\mu_p}|0\rangle
\end{eqnarray}
as the covariant derivative operator
\begin{eqnarray}
D_\mu|\varphi\rangle
&=&
\sum_{p=0}^{}(\nabla_\mu\varphi_{\mu_1\ldots\mu_p})\;
a^{\mu_1+}\ldots a^{\mu_p}|0\rangle.
\end{eqnarray}

Now we want to realize equations (\ref{EM-curved}) (with $m=0$) as operator
constraints in the Fock space.
For this purpose let us define operators
\begin{eqnarray}
l_0=D^2+X,
&\qquad&
l_1=-ia^\mu D_\mu
\end{eqnarray}
where
$D^2=g^{\mu\nu}(D_\mu D_\nu-\Gamma_{\mu\nu}^\sigma D_\sigma)$
and the operator $X$ is responsible for the ``terms with
curvature''
in the first equation of (\ref{EM-curved}).
Then the equations
\begin{eqnarray}
l_0|\varphi\rangle=0,
&\qquad&
l_1|\varphi\rangle=0
\label{EM-op-0}
\end{eqnarray}
are equivalent to the corresponding equations in
(\ref{EM-curved}).

In order to construct Lagrangian within the BRST approach we
must have at hand a set of operators which is invariant under
Hermitian conjugation and which forms an algebra
\cite{massive-bos}.
We assume the standard scalar product in the Fock space and
suppose that operator $X$ and hence operator $l_0$ are hermitian
with respect to this scalar product.
The operator conjugate to $l_1$ we denote as $l_1^+$
\begin{eqnarray}
l_1^+=-ia^{\mu+}D_\mu.
\end{eqnarray}
Now set of operators $l_0$, $l_1$, $l_1^+$ is invariant under
Hermitian conjugation. Then we must realize the second
requirement: one should obtain set of operators which form an
algebra. For this purpose we find all (anti)commutators
generated by $l_0$, $l_1$, $l_1^+$. Since operator $l_0$ is not
yet defined we calculate the anticommutator $\{l_1,l_1^+\}$.
One has
\begin{eqnarray}
\{l_1,l_1^+\}&=&
-D^2-R_{\mu\nu\alpha\beta}\;a^{+\mu} a^\nu a^{+\alpha}a^\beta
\label{l1l1+}
\end{eqnarray}
where
\begin{math}
R^\alpha{}_{\beta\mu\nu}=
\partial_\mu \Gamma_{\nu\beta}^\alpha-\partial_\nu \Gamma_{\mu\beta}^\alpha
+\Gamma_{\mu\rho}^\alpha\Gamma_{\nu\beta}^\rho
-\Gamma_{\nu\rho}^\alpha\Gamma_{\mu\beta}^\rho
.
\end{math}
Since the rhs of
(\ref{l1l1+}) contains operator $D^2$ which are present in
operator $l_0$  we rewrite (\ref{l1l1+}) as follows
\begin{eqnarray}
\{l_1,\l_1^+\}&=&
-l_0+X-R_{\mu\nu\alpha\beta}\;a^{+\mu} a^\nu a^{+\alpha}a^\beta
.
\label{X}
\end{eqnarray}
From (\ref{X}) we see that in order  to close the algebra we must put
$X=R_{\mu\nu\alpha\beta}\;a^{+\mu} a^\nu a^{+\alpha}a^\beta$
and as a consequence we have operator $l_0$ in the form
\begin{eqnarray}
l_0=
D^2+R_{\mu\nu\alpha\beta}\;a^{+\mu} a^\nu a^{+\alpha}a^\beta
.
\label{l0}
\end{eqnarray}
One can check that set of operators $l_0$, $l_1$, $l_1^+$ form
an algebra
\begin{eqnarray}
\{l_1,l_1^+\}=-l_0,
\label{alg-0}
&\qquad&
\{l_1,l_1\}=\{l_1^+,l_1^+\}=[l_1,l_0]=[l_1^+,l_0]=0.
\end{eqnarray}
Thus now we have at hand set of operators which is invariant
under Hermitian conjugation and form an algebra.
Let us note that found expression for operator $l_0$ gives the
following mass-shell equation on antisymmetric field of rank-$p$
in arbitray curved space
\begin{eqnarray}
\nabla^2\varphi_{\mu_1\ldots\mu_p}
+(-1)^ppR_{[\mu_1}^\alpha\varphi_{\mu_2\ldots\mu_p]\alpha}
-p(p-1)R_{[\mu_1}{}^{\alpha\beta}{}_{\mu_2}\varphi_{\mu_3\ldots\mu_p]\alpha\beta}
=0
\label{msh-0}
\end{eqnarray}
where the square brackets denote antisymmetrization
\begin{eqnarray}
A_{[\alpha_1\ldots\alpha_p]}&=&\frac{1}{p!}
\Bigl[  A_{\alpha_1\ldots\alpha_p}\pm (p!-1)
\mbox{\it{}permutations}\Bigr]
.
\end{eqnarray}

Let us turn to construction of the Lagrangians.
Among the operators of the algebra (\ref{alg-0}) there are no
operators which are not constraints. All the operators are
constraints in the bra-vector space or/and in the ket-vector
space
\begin{eqnarray}
\langle\varphi|l_0=\langle\varphi|l_1^+=0,
&\qquad&
l_0|\varphi\rangle=l_1|\varphi\rangle=0.
\end{eqnarray}
Therefore for constructing Lagrangians there is no need to introduce enlarged expressions for
the operators \cite{massive-bos}
and one may construct BRST operator from the operators $l_0$,
$l_1$, $l_1^+$. For this we introduce ghost `coordinates'
$\eta_0$, $q_1^+$, $q_1$ and canonically conjugated them
`momenta' $\mathcal{P}_0$, $p_1$, $p_1^+$ with nonvanishing
(anti)commutators
\begin{eqnarray}
\label{bghosts}
&&
\{\eta_0,\mathcal{P}_0\}=1,
\qquad
[q_1,p_1^+]=[q_1^+,p_1]=i.
\end{eqnarray}
After this one finds the BRST operator
\begin{eqnarray}
Q
&=&
\eta_0l_{0}+q_1^+l_1+q_1l_1^+
+q_1^+q_1{\cal{}P}_0
,
\hspace*{4.5em}
Q^2=0.
\label{Q-0}
\end{eqnarray}
Further we define the representation of the Hilbert space where
the BRST operators acts as follows
\begin{eqnarray}
\label{bghosts0}
a^\mu|0\rangle=q_1|0\rangle=p_1|0\rangle=\mathcal{P}_0|0\rangle=0
\end{eqnarray}
and as a consequence the general form of the state vector in the
Hilbert space is
\begin{eqnarray}
|\Phi\rangle&=&\sum_{k_i}\eta_0^{k_1}(q_1^+)^{k_2}(p_1^+)^{k_3}
a^{+\mu_1}\ldots a^{+\mu_{k_0}}
\Phi_{\mu_1\ldots{}\mu_{k_0}}^{k_1k_2k_3}(x)|0\rangle.
\label{GState-0}
\end{eqnarray}
The sum in (\ref{GState-0}) is taken over $k_1$ running from 0 to 1
and over $k_2$, $k_3$, $k_0$ running from 0 to infinity.

Let us introduce operator
\begin{eqnarray}
&&
\sigma_0
=
a_\mu^+a^\mu+iq_1^+p_1-ip_1^+q_1
,
\hspace*{5em}
[Q,\sigma_0]=0
,
\end{eqnarray}
which commute with the BRST operator.
This operator is used for constructing Lagrangian with given
spin $p$. For this we restrict the fields $|\Phi\rangle$ and the
gauge parameters $|\Lambda^{(i)}\rangle$ in the extended Fock
space including ghosts (\ref{GState-0}) by the conditions
\begin{eqnarray}
\sigma_0|\Phi\rangle=p|\Phi\rangle,
&\qquad&
\sigma_0|\Lambda^{(i)}\rangle=p|\Lambda^{(i)}\rangle.
\label{sigma}
\end{eqnarray}
If we omit these conditions then the Lagrangian (and the gauge
transformations) will contain fields with all spins
simultaneously.
One can show (see e.g. \cite{massive-bos})
that Lagrangian can be written as
\begin{eqnarray}
\mathcal{L}&=&\int d\eta_0\, \langle\Phi|Q|\Phi\rangle,
\label{L0}
\end{eqnarray}
which is invariant under the reducible gauge transformations
\begin{eqnarray}
\delta|\Phi\rangle=Q|\Lambda^{(0)}\rangle,
\quad\ldots\quad
\delta|\Lambda^{(i)}\rangle=Q|\Lambda^{(i+1)}\rangle,
\quad\ldots\quad
\delta|\Lambda^{(p-2)}\rangle=Q|\Lambda^{(p-1)}\rangle.
\label{GT0}
\end{eqnarray}
The chain of the gauge transformation for each given spin $p$ is finite due to
(\ref{sigma}) and to the ghost number restriction
\begin{eqnarray}
gh(|\Phi\rangle)=0,
&\qquad&
gh(|\Lambda^{(i)}\rangle=-(i+1).
\label{gh0}
\end{eqnarray}
Thus the Lagrangian for the massless bosonic antisymmetric field
in an arbitrary curved background
is constructed.

Let us show that Lagrangian (\ref{L0}) reproduces equations of
motion (\ref{EM-curved}), (\ref{msh-0}) [or the same
(\ref{EM-op-0})] after gauge fixing. Let
us fix the rank of the antisymmetric field to be $p$. In this
case we have $p-1$ reducibility stages of the gauge symmetry and
due to
(\ref{sigma}) and the ghost number restriction (\ref{gh0})
the lowest stage gauge parameter $|\Lambda^{(p-1)}\rangle$ can not depend on ghost
$\eta_0$: $\mathcal{P}_0|\Lambda^{(p-1)}\rangle=0$.
Then we introduce the following decomposition on ghost $\eta_0$ of
the gauge parameters
\begin{eqnarray}
|\Lambda^{(i)}\rangle=|\Lambda^{(i)}_0\rangle+\eta_0|\Lambda^{(i)}_1\rangle.
\end{eqnarray}
One can show that using gauge transformation for the $p-2$
stage gauge parameter
\begin{eqnarray}
\delta|\Lambda^{(p-2)}_0\rangle=(q_1^+l_1+q_1l_1^+)|\Lambda^{(p-1)}_0\rangle,
&\qquad&
\delta|\Lambda^{(p-2)}_1\rangle=l_0|\Lambda^{(p-1)}_0\rangle
\end{eqnarray}
we can get rid of the dependence of $p-2$ stage gauge parameter
on ghost $\eta_0$. Then we can repeat the procedure and get rid
of the dependence of $p-3$ stage gauge parameter on ghost
$\eta_0$ using remaining part $|\Lambda^{(p-2)}_0\rangle$ of the
gauge parameter $|\Lambda^{(p-2)}\rangle$. Applying the same
procedure further we remove dependence of the gauge parameter
$|\Lambda^{(0)}\rangle$ on ghost $\eta_0$. Thus it remains only part of
the gauge parameter
$|\Lambda^{(0)}\rangle$ which independent of $\eta_0$:
$|\Lambda^{(0)}_0\rangle$.
Now we decompose field $|\Phi\rangle$ and gauge parameter
$|\Lambda^{(0)}_0\rangle$
 satisfying (\ref{gh0}) and (\ref{sigma}) with given
$p$ as follows\footnote{In decomposition (\ref{Phi-0}) the
physical field $\varphi(x)_{\mu_1\ldots\mu_p}$ is contained in
$|\varphi_p\rangle$. All other fields $|\varphi_n\rangle$ are auxiliary ones.}
\begin{eqnarray}
|\Phi\rangle&=&
\sum_{n=0}^{[p/2]}\frac{(-iq_1^+p_1^+)^n}{n!}|\varphi_{p-2n}\rangle
+\eta_0
\sum_{n=1}^{[(p+1)/2]}(q_1^+)^{n-1}\frac{(-ip_1^+)^n}{n!}|\varphi_{p-2n+1}\rangle
,
\label{Phi-0}
\\
|\Lambda^{(0)}_0\rangle&=&
\sum_{n=1}^{[(p+1)/2]}(q_1^+)^{n-1}\frac{(-ip_1^+)^n}{n!}|\lambda_{p-2n+1}\rangle
,
\end{eqnarray}
where we denote
\begin{math}
|\varphi_n\rangle=
\frac{(-i)^n}{n!}
\varphi(x)_{\mu_1\ldots\mu_n}a^{+\mu_1}\ldots
a^{+\mu_n}|0\rangle
\end{math}
and
\begin{math}
|\lambda_n\rangle=
\frac{(-i)^n}{n!}
\lambda(x)_{\mu_1\ldots\mu_n}a^{+\mu_1}\ldots
a^{+\mu_n}|0\rangle
.
\end{math}
Equation of motion $Q|\Phi\rangle=0$ and gauge transformation
$\delta|\Phi\rangle=Q|\Lambda^{(0)}_0\rangle$ looks like
\begin{eqnarray}
&&
l_0|\varphi_{p-2n}\rangle=l_1^+|\varphi_{p-2n-1}\rangle
+l_1|\varphi_{p-2n+1}\rangle,
\label{EM-0}
\\
&&
|\varphi_{p-2n-1}\rangle+l_1|\varphi_{p-2n}\rangle+l_1^+|\varphi_{p-2n-2}\rangle=0,
\label{EM-1}
\\[0.5em]
&&
\delta|\varphi_{p-2n}\rangle=l_1|\lambda_{p-2n+1}\rangle+l_1^+|\lambda_{p-2n-1}\rangle,
\label{d2n}
\\
&&
\delta|\varphi_{p-2n-1}\rangle=l_0|\lambda_{p-2n-1}\rangle.
\label{d2n-1}
\end{eqnarray}
Using gauge transformation (\ref{d2n}) starting with the field
with the lowest rank we eliminate all the fields
$|\varphi_{p-2n}\rangle$ except $|\varphi_p\rangle$.
After this all the fields $|\varphi_{p-2n-1}\rangle$ except
$|\varphi_{p-1}\rangle$ become zero as consequences equations of
motion (\ref{EM-1}).
The rest equations of motion on fields $|\varphi_p\rangle$ and
$|\varphi_{p-1}\rangle$ and the residual gauge transformation
with restricted gauge parameter $|\lambda_{p-1}\rangle$
are
\begin{align}
&
l_0|\varphi_p\rangle=l_1^+|\varphi_{p-1}\rangle,
&&
\delta|\varphi_p\rangle=l_1^+|\lambda_{p-1}\rangle,
\\
&
l_1|\varphi_p\rangle+|\varphi_{p-1}\rangle=0,
\label{lfi1}
&&
\delta|\varphi_{p-1}\rangle=l_0|\lambda_{p-1}\rangle,
&&
l_1|\lambda_{p-1}\rangle=0.
\end{align}
Acting on equation of motion (\ref{lfi1}) by operator $l_1$ and
using that $l_1^2=0$ (\ref{alg-0}) one finds that
$l_1|\varphi_{p-1}\rangle=0$.
Therefore using the residual gauge transformation with
restricted gauge parameter $|\lambda_{p-1}\rangle$ we can make
solution $|\varphi_{p-1}\rangle$ to be zero
$|\varphi_{p-1}\rangle=0$.
As a result only physical field $|\varphi_p\rangle$ is nonvanishing
 and equations of motion for it are
\begin{eqnarray}
l_0|\varphi_p\rangle=0,
&\qquad&
l_1|\varphi_p\rangle=0.
\end{eqnarray}
Thus we have shown that equations of motion following from
Lagrangian (\ref{L0}) gives (\ref{EM-curved}), (\ref{msh-0}) [or
equivalently (\ref{EM-op-0})] up to a gauge fixing.

Let us simplify Lagrangian (\ref{L0}) for antisymmetric field with given
rank-$p$.
Substituting decomposition (\ref{Phi-0}) of the field
$|\Phi\rangle$ into (\ref{L0}) one gets
\begin{eqnarray}
\mathcal{L}&=&
\langle\varphi_p|\bigl\{l_0|\varphi_p\rangle-l_1^+|\varphi_{p-1}\rangle\bigr\}
+\sum_{n=1}^{[p/2]-1}\langle\varphi_{p-2n}|\bigl\{l_0|\varphi_{p-2n}\rangle
 -l_1|\varphi_{p-2n+1}\rangle-l_1^+|\varphi_{p-2n-1}\rangle\bigr\}
\nonumber
\\
&&{}
-\sum_{n=0}^{[(p-3)/2]}\langle\varphi_{p-2n-1}|\bigl\{|\varphi_{p-2n-1}\rangle
   +l_1|\varphi_{p-2n}\rangle+l_1^+|\varphi_{p-2n-2}\rangle\bigr\}
\nonumber
\\
&&{}
+
\left\{
\begin{array}{ll}
p~even&
\langle\varphi_0|\bigl\{l_0|\varphi_0\rangle-l_1|\varphi_1\rangle\bigr\}
-
\langle\varphi_1|\bigl\{|\varphi_1\rangle+l_1|\varphi_2\rangle
   +l_1^+|\varphi_0\rangle\bigr\}
\\[1em]
p~odd&
\langle\varphi_1|\bigl\{l_0|\varphi_1\rangle-l_1|\varphi_2\rangle
    -l_1^+|\varphi_0\rangle\bigr\}
-
\langle\varphi_0|\bigl\{|\varphi_0\rangle+l_1|\varphi_1\rangle\bigr\}
\end{array}
\right.
\label{Lf}
\end{eqnarray}
Here field $|\varphi_p\rangle$ is a physical one and the rest
fields are auxiliary. Our purpose now is to obtain Lagrangian in
terms of one physical field $|\varphi_p\rangle$.
For this we express fields $|\varphi_{p-2n+1}\rangle$ using
their equations of motion
\begin{math}
|\varphi_{p-2n+1}\rangle
=
-l_1|\varphi_{p-2n+2}\rangle
-l_1^+|\varphi_{p-2n}\rangle
\end{math}
and substitute them into Lagrangian (\ref{Lf}).
Taking into account (\ref{alg-0}) one can show that all other
fields except $|\varphi_p\rangle$ disappear and we get
\begin{eqnarray}
\mathcal{L}&=&
\langle\varphi_p|[l_0+l_1^+l_1]|\varphi_p\rangle
=
-\langle\varphi_p|l_1l_1^+|\varphi_p\rangle
.
\end{eqnarray}
In order to obtain this Lagrangian in component form we substitute
into it the explicit expressions for $l_1$, $l_1^+$, and for
\begin{math}
|\varphi_p\rangle=
\frac{(-i)^p}{p!}
\varphi(x)_{\mu_1\ldots\mu_n}a^{+\mu_1}\ldots
a^{+\mu_p}|0\rangle
\end{math}
and obtain
\begin{eqnarray}
\mathcal{L}&=&-\frac{1}{(p+1)!}\;
F_{\mu_1\ldots\mu_{p+1}}F^{\mu_1\ldots\mu_{p+1}}
\label{L0f}
\end{eqnarray}
where $F_{\mu_1\ldots\mu_{p+1}}$ is the strength of
antisymmetric field $\varphi_{\mu_1\ldots\mu_p}$
\begin{eqnarray}
F_{\mu_1\ldots\mu_{p+1}}&=&
\frac{1}{p!}\Bigl[
\nabla_{\mu_1}\varphi_{\mu_2\ldots\mu_{p+1}}
\pm (p+1)!-1~permutations
\Bigr]
.
\label{strength}
\end{eqnarray}
Thus we simplified Lagrangian (\ref{L0}) and obtained Lagrangian
in terms of one physical field.
It is this form of Lagrangian which is commonly used for
antisymmetric massless field.

Let us turn to Lagrangian construction for massive antisymmetric
fields.

\section{Lagrangian construction for massive fields}

In the massive case we take the mass-shell equation in the form
of the mass-shell equation for the massless
antisymmetric field (\ref{msh-0}) deformed by the mass term
\begin{eqnarray}
(\nabla^2-m^2)\varphi_{\mu_1\ldots\mu_p}
+(-1)^ppR_{[\mu_1}^\alpha\varphi_{\mu_2\ldots\mu_p]\alpha}
-p(p-1)R_{[\mu_1}{}^{\alpha\beta}{}_{\mu_2}\varphi_{\mu_3\ldots\mu_p]\alpha\beta}
=0.
\label{msh-m}
\end{eqnarray}
Then
we denote operator corresponding to
equation (\ref{msh-m}) as $l_0^{(m)}$ and define it as
$l_0^{(m)}=l_0-m^2$
\begin{eqnarray}
l_0^{(m)}=
D^2-m^2+R_{\mu\nu\alpha\beta}\;a^{+\mu} a^\nu a^{+\alpha}a^\beta
.
\end{eqnarray}
Now anticommutator $\{l_1,l_1^+\}$ is
\begin{eqnarray}
\{l_1,l_1^+\}&=&-l_0^{(m)}-m^2
\end{eqnarray}
and in order to have a set of operators which is invariant
under Hermitian conjugation and which form an algebra we add
operator $g_m=m^2$.
As a result the algebra of the operators in the massive case is
\begin{eqnarray}
\{l_1,l_1^+\}=-l_0^{(m)}-g_m,
&\qquad&
\{l_1,l_1\}=\{l_1^+,l_1^+\}=[l_1,l_0^{(m)}]=[l_1^+,l_0^{(m)}]=0,
\\
&&
[g_m,l_1]=[g_m,l_1^+]=[g_m,l_0]=0.
\end{eqnarray}
In the set of operators we have one operator $g_m$ which is not
a constraint neither in the bra nor in the ket-vector space.
In this case in order to construct Lagrangian within the BRST approach
(see e.g. \cite{massive-bos})
we
need to introduce additional (new) creation and annihilation
operators and then construct extended operators
$o_i\to{}O_i=o_i+o_i'$, $o_i=(l_0^{(m)},l_1,l_1^+,g_m)$
which must satisfy two conditions: 1)~they must form an algebra
$[O_i,O_j]\sim{}O_k$;
2)~the operators which are not constraints must be zero (that is
in the case under consideration we must have $G_m=g_m+g_m'=0$).

For this purpose we introduce one pair fermionic creation and
annihilation operators with the standard commutation relations
$\{f,f^+\}=1$ and put
\begin{align}
&
l_0^{(m)\prime}=0,
&&
l_1^{\prime+}=mf^+,
&&
l_1'=mf,
&&
g_m'=-m^2.
\end{align}
One can check that the extended expressions of the operators satisfy the above
requirements: the extended expression for the operator $g_m$
which is not a constraint is zero\footnote{In what follows we
forget about operator $G_m=0$.} $G_m=0$; the operators $L_0^{(m)}$,
$L_1$, $L_1^+$ form an algebra
\begin{eqnarray}
\{L_1,L_1^+\}=-L_0^{(m)},
&\qquad&
\{L_1,L_1\}=\{L_1^+,L_1^+\}=[L_1,L_0^{(m)}]=[L_1^+,L_0^{(m)}]=0.
\end{eqnarray}
After this one should construct BRST operator
\begin{eqnarray}
Q_m
&=&
\eta_0L_0^{(m)}+q_1^+L_1+q_1L_1^+
+q_1^+q_1{\cal{}P}_0
,
\hspace*{6.2em}
Q_m^2=0,
\label{Q-m}
\\
&&
\sigma_m
=
a_\mu^+a^\mu+f^+f+iq_1^+p_1-ip_1^+q_1
,
\hspace*{5em}
[Q_m,\sigma_m]=0
,
\label{sigma-m}
\end{eqnarray}
with the ghosts satisfying relations (\ref{bghosts}), (\ref{bghosts0}).
In the massive case the general state in the Hilbert space looks
as follows
\begin{eqnarray}
|\Phi\rangle&=&\sum_{k_i}\eta_0^{k_1}(q_1^+)^{k_2}(p_1^+)^{k_3}
(f^+)^{k_4}a^{+\mu_1}\ldots a^{+\mu_{k_0}}
\Phi_{\mu_1\ldots{}\mu_{k_0}}^{k_1k_2k_3k_4}(x)|0\rangle.
\label{GState-m}
\end{eqnarray}
The sum in (\ref{GState-m}) is taken over $k_1$, $k_4$ running from 0 to 1
and over $k_2$, $k_3$, $k_0$ running from 0 to infinity.

Analogously to the massless case to construct Lagrangian for a field
with a given spin $p$ we restrict the fields $|\Phi\rangle$ and the
gauge parameters $|\Lambda^{(i)}\rangle$ in the extended Fock space
(\ref{GState-m}) as follows
\begin{eqnarray}
\sigma_m|\Phi\rangle=p|\Phi\rangle,
&\qquad&
\sigma_m|\Lambda^{(i)}\rangle=p|\Lambda^{(i)}\rangle,
\label{sigma-bm}
\end{eqnarray}
with operator $\sigma_m$ given in (\ref{sigma-m})
If we omit these conditions then the Lagrangian (and the gauge
transformations) will contain fields with all spins.
One can show (see e.g. \cite{massive-bos})
that Lagrangian can be written as
\begin{eqnarray}
\mathcal{L}&=&\int d\eta_0\, \langle\Phi|Q_m|\Phi\rangle,
\label{Lm}
\end{eqnarray}
which is invariant under the reducible gauge transformations
\begin{eqnarray}
\label{GTm}
\delta|\Phi\rangle=Q_m|\Lambda^{(0)}\rangle,
\quad\ldots\quad
\delta|\Lambda^{(i)}\rangle=Q_m|\Lambda^{(i+1)}\rangle,
\quad\ldots\quad
\delta|\Lambda^{(p-2)}\rangle=Q_m|\Lambda^{(p-1)}\rangle.
\end{eqnarray}
We note here that in the massive case the gauge symmetry are St\"uckelberg
one. The St\"uckelberg fields and gauge parameters in
decomposition (\ref{GState-m}) are those which contain $f^+$,
i.e. fields corresponding to $k_4=1$:
$\Phi_{\mu_1\ldots{}\mu_{k_0}}^{k_1k_2k_31}(x)$.
The chain of the gauge transformations (\ref{GTm}) is finite due to
(\ref{sigma-bm}) and to the ghost number restriction
\begin{eqnarray}
gh(|\Phi\rangle)=0,
&\qquad&
gh(|\Lambda^{(i)}\rangle=-(i+1).
\label{gh-m}
\end{eqnarray}
Thus the Lagrangian for the massive bosonic antisymmetric field
in an arbitrary curved background
is constructed.

Let us show that Lagrangian (\ref{Lm}) gives equations of
motion which equivalent (\ref{EM-curved}), (\ref{msh-m}) up to a gauge transformation.
Let the rank of the antisymmetric field be equal to $p$. Then
we have a reducible gauge theory with $p-1$ number of
reducibility stages. Due to (\ref{sigma-bm}) and (\ref{gh-m})
the lowest stage gauge parameter has the form
\begin{eqnarray}
|\Lambda^{(p-1)}\rangle&=&
\frac{(-i)^p}{p!}(p_1^+)^p\xi^{(p-1)}(x)|0\rangle
.
\end{eqnarray}
It can be explicitly checked that we can eliminate the dependence on
$f^+$ in the gauge function $|\Lambda^{(p-2)}\rangle$ of the
$(p-2)$-stage. Then it is possible to check that we can remove
dependence of $|\Lambda^{(p-3)}\rangle$ on $f^+$ with the help of
the remaining gauge parameters $|\Lambda^{(p-2)}\rangle$ (which do
not depend on $f^+$). Then we can proceed further in the same way
and in the end remove dependence of the field $|\Phi\rangle$ on
$f^+$. Thus we have the gauge $f|\Phi\rangle=0$ and all the gauge
parameters have been used.

Let us turn to the equations of motion $Q_m|\Phi\rangle=0$.
Decomposing in power series of $f^+$ one finds that a
part of the equations of motion has the form
\begin{eqnarray}
mq_1f^+|\Phi\rangle=0,
\end{eqnarray}
which tells us that the field does not depend on $p_1^+$. This
condition together with the gauge $f|\Phi\rangle=0$ give that
the field $|\Phi\rangle$ can depend on $a^{+\mu}$ only.
That is it remains only the physical field
\begin{eqnarray}
|\Phi\rangle
=|\varphi_p\rangle
=\frac{(-i)^p}{p!}
\varphi(x)_{\mu_1\ldots\mu_p}a^{+\mu_1}\ldots
a^{+\mu_p}|0\rangle
\end{eqnarray}
and the equations of motion for it are
\begin{eqnarray}
l_0^{(m)}|\varphi_p\rangle=0,
&\qquad&
l_1|\varphi_p\rangle=0.
\end{eqnarray}
In component form these are equations (\ref{EM-curved}),
(\ref{msh-m}) which we wanted to reproduce.

Let us simplify Lagrangian (\ref{Lm}). For this purpose we
decompose field $|\Phi\rangle$ satisfying (\ref{sigma-bm}) with
a given $p$ and (\ref{gh-m}) as follows\footnote{Analogously to
massless case the
physical field $\varphi(x)_{\mu_1\ldots\mu_p}$ in decomposition
(\ref{Phi-m}) is contained in
$|\varphi_p\rangle$. The other fields $|\varphi_n\rangle$ are
auxiliary and fields $|\psi_n\rangle$ are St\"uckelberg ones.}
\begin{eqnarray}
|\Phi\rangle&=&
\sum_{n=0}^{[p/2]}\frac{(-iq_1^+p_1^+)^n}{n!}|\varphi_{p-2n}\rangle
+\eta_0\sum_{n=1}^{[(p+1)/2]}(q_1^+)^{n-1}\frac{(-ip_1^+)^n}{n!}|\varphi_{p-2n+1}\rangle
\nonumber
\\
&&{}
+f^+\sum_{n=0}^{[(p-1)/2]}\frac{(-iq_1^+p_1^+)^n}{n!}|\psi_{p-2n-1}\rangle
+\eta_0f^+\sum_{n=1}^{[p/2]}(q_1^+)^{n-1}\frac{(-ip_1^+)^n}{n!}|\psi_{p-2n}\rangle
\label{Phi-m}
\end{eqnarray}
where as before we denote
\begin{eqnarray}
|\varphi_n\rangle=\frac{(-i)^n}{n!}\varphi(x)_{\mu_1\ldots\mu_n}
a^{+\mu_1}\ldots a^{+\mu_n}|0\rangle,
\qquad
|\psi_n\rangle=\frac{(-i)^n}{n!}\psi(x)_{\mu_1\ldots\mu_n}
a^{+\mu_1}\ldots a^{+\mu_n}|0\rangle.
\end{eqnarray}
Then we substitute (\ref{Phi-m}) into (\ref{Lm}) and obtain
\begin{eqnarray}
\mathcal{L}&=&
\langle\varphi_p|\bigl\{l_0^{(m)}|\varphi_p\rangle-l_1^+|\varphi_{p-1}\rangle\bigr\}
+
\langle\psi_{p-1}|\bigl\{l_0^{(m)}|\psi_{p-1}\rangle-m|\varphi_{p-1}\rangle
+l_1^+|\psi_{p-2}\rangle\bigr\}
\nonumber
\\
&&{}
+\sum_{n=1}^{[p/2]-1}\langle\varphi_{p-2n}|\bigl\{l_0^{(m)}|\varphi_{p-2n}\rangle
 -l_1|\varphi_{p-2n+1}\rangle-l_1^+|\varphi_{p-2n-1}\rangle-m|\psi_{p-2n}\rangle\bigr\}
\nonumber
\\
&&{}
-\sum_{n=0}^{[(p-3)/2]}\langle\varphi_{p-2n-1}|\bigl\{|\varphi_{p-2n-1}\rangle
   +l_1|\varphi_{p-2n}\rangle+l_1^+|\varphi_{p-2n-2}\rangle+m|\psi_{p-2n-1}\rangle\bigr\}
\nonumber
\\
&&{}
+\sum_{n=1}^{[(p-3)/2]}\langle\psi_{p-2n-1}|\bigl\{
l_0^{(m)}|\psi_{p-2n-1}\rangle-m|\varphi_{p-2n-1}\rangle
+l_1|\psi_{p-2n}\rangle+l_1^+|\psi_{p-2n-2}\rangle
\bigr\}
\nonumber
\\
&&{}
-\sum_{n=1}^{[p/2]-1}\langle\psi_{p-2n}|\bigl\{|\psi_{p-2n}\rangle
+m|\varphi_{p-2n}\rangle-l_1|\psi_{p-2n+1}\rangle
-l_1^+|\psi_{p-2n-1}\rangle\bigr\}
\nonumber
\\
&&{}
+
\left\{
\begin{array}{ll}
p~even&
\langle\varphi_0|\bigl\{l_0^{(m)}|\varphi_0\rangle
   -l_1|\varphi_1\rangle-m|\psi_0\rangle\bigr\}
-
\langle\varphi_1|\bigl\{|\varphi_1\rangle+l_1|\varphi_2\rangle
   +l_1^+|\varphi_0\rangle+m|\psi_1\rangle\bigr\}
\\[1em]
p~odd&
\langle\varphi_1|\bigl\{l_0^{(m)}|\varphi_1\rangle-l_1|\varphi_2\rangle
    -l_1^+|\varphi_0\rangle-m|\psi_1\rangle\bigr\}
-
\langle\varphi_0|\bigl\{|\varphi_0\rangle+l_1|\varphi_1\rangle+m|\psi_0\rangle\bigr\}
\end{array}
\right.
\nonumber
\\
&&{}
+
\left\{
\begin{array}{ll}
p~even&
\langle\psi_1|\bigl\{l_0^{(m)}|\psi_1\rangle-m|\varphi_1\rangle
+l_1|\psi_2\rangle+l_1^+|\psi_0\rangle\bigr\}
-
\langle\psi_0|\bigl\{|\psi_0\rangle+m|\varphi_0\rangle-l_1|\psi_1\rangle\bigr\}
\\[1em]
p~odd&
\langle\psi_0|\bigl\{l_0^{(m)}|\psi_0\rangle-m|\varphi_0\rangle
+l_1|\psi_1\rangle\bigr\}
-
\langle\psi_1|\bigl\{|\psi_1\rangle+m|\varphi_1\rangle
-l_1|\psi_2\rangle-l_1^+|\psi_0\rangle\bigr\}
\end{array}
\right.
\label{Lfm}
\end{eqnarray}
After this we substitute equations of motion
\begin{math}
|\varphi_{p-2n-1}\rangle=-l_1|\varphi_{p-2n}\rangle
-l_1^+|\varphi_{p-2n-2}\rangle-m|\psi_{p-2n-1}\rangle
\end{math}
and
\begin{math}
|\psi_{p-2n}\rangle=l_1|\psi_{p-2n+1}\rangle
+l_1^+|\psi_{p-2n-1}\rangle-m|\varphi_{p-2n}\rangle
\end{math}
into Lagrangian (\ref{Lfm}) and find
\begin{eqnarray}
\mathcal{L}&=&
\langle\varphi_p|\bigl\{[l_0^{(m)}+l_1^+l_1]|\varphi_p\rangle
+ml_1^+|\psi_{p-1}\rangle\bigr\}
+
\langle\psi_{p-1}|\bigl\{[l_0^{(m)}+m^2+l_1^+l_1]|\psi_{p-1}\rangle
+ml_1|\varphi_p\rangle\bigr\}
\nonumber
\\[0.5em]
&=&
-\langle\varphi_p|[l_1l_1^++m^2]|\varphi_p\rangle
-\langle\psi_{p-1}|l_1l_1^+|\psi_{p-1}\rangle
+\langle\varphi_p|ml_1^+|\psi_{p-1}\rangle
+\langle\psi_{p-1}|ml_1|\varphi_p\rangle
.
\label{LmSt}
\end{eqnarray}
In order to get gauge transformations which leave this
Lagrangian invariant we decompose gauge parameters
$|\Lambda^{(i)}\rangle$ satisfying (\ref{gh-m}) and
(\ref{sigma-bm}) with given $p$ as follows
\begin{eqnarray}
|\Lambda^{(i)}\rangle&=&
\sum_{n=0}^{[(p-i-1)/2]}(q_1^+)^{n}
  \frac{(-ip_1^+)^{n+i+1}}{(n+i+1)!}|\lambda^{(i)}_{p-2n-i-1}\rangle
+\eta_0\sum_{n=0}^{[(p-i)/2]}(q_1^+)^{n}
  \frac{(-ip_1^+)^{n+i+2}}{(n+i+2)!}|\lambda^{(i)}_{p-2n-i}\rangle
\\
&&{}
+f^+\sum_{n=0}^{[(p-i-2)/2]}(q_1^+)^{n}
  \frac{(-ip_1^+)^{n+i+1}}{(n+i+1)!}|\lambda^{(i)}_{p-2n-i-2}\rangle
+\eta_0f^+\sum_{n=0}^{[(p-i-1)/2]}(q_1^+)^{n}
  \frac{(-ip_1^+)^{n+i+2}}{(n+i+2)!}|\lambda^{(i)}_{p-2n-i-1}\rangle
.
\nonumber
\end{eqnarray}
Substituting this decomposition into (\ref{GTm}) one finds the
chain of reducible gauge transformation under which (\ref{LmSt}) is
invariant
\begin{align}
&
\delta|\varphi_p\rangle=l_1^+|\lambda_{p-1}\rangle,
&&
\delta|\psi_{p-1}\rangle=m|\lambda_{p-1}\rangle+l_1^+|\omega_{p-2}\rangle
\\
&\ldots&&\ldots
\nonumber
\\
&
\delta|\lambda^{(i)}_{p-i-1}\rangle=l_1^+|\lambda^{(i+1)}_{p-i-2}\rangle,
&&
\delta|\omega^{(i)}_{p-i-2}\rangle=m|\lambda^{(i+1)}_{p-i-2}\rangle
+l_1^+|\omega^{(i+1)}_{p-i-3}\rangle,
\\
&\ldots&&\ldots
\nonumber
\\
&
\delta|\lambda^{(p-2)}_1\rangle=l_1^+|\lambda^{(p-1)}_0\rangle,
&&
\delta|\omega^{(p-2)}_0\rangle=m|\lambda^{(p-1)}_0\rangle.
\end{align}
We see that field $|\psi_{p-1}\rangle$ and gauge parameters
$|\omega^{(i)}_{p-i-2}\rangle$ are responsible for the presence
of the St\"uckelberg symmetry. We can partially fix the gauge
symmetry and get rid of the St\"uckelberg gauge parameters
$|\omega^{(i)}_{p-i-2}\rangle$. After this Lagrangian
(\ref{LmSt}) will be invariant under the following St\"uckelberg symmetry transformation
\begin{align}
&
\delta|\varphi_p\rangle=l_1^+|\lambda_{p-1}\rangle,
&&
\delta|\psi_{p-1}\rangle=m|\lambda_{p-1}\rangle.
\label{gtm}
\end{align}
Let us write Lagrangian (\ref{LmSt}) and gauge transformations
(\ref{gtm}) in component form. One has
\begin{eqnarray}
\mathcal{L}&=&
-\frac{1}{(p+1)!}\;
F_{\mu_1\ldots\mu_{p+1}}F^{\mu_1\ldots\mu_{p+1}}
-\frac{m^2}{p!}\;\varphi_{\mu_1\ldots\mu_p}\varphi^{\mu_1\ldots\mu_p}
\nonumber
\\
&&\hspace*{14em}{}
-\frac{1}{p!}\;f_{\mu_1\ldots\mu_{p}}f^{\mu_1\ldots\mu_{p}}
+\frac{2m}{p!}\;\varphi_{\mu_1\ldots\mu_p}f^{\mu_1\ldots\mu_p}
\label{Lm-1}
\end{eqnarray}
\begin{eqnarray}
&&
\delta\varphi_{\mu_1\ldots\mu_p}=p\;\nabla_{[\mu_1}\lambda_{\mu_2\ldots\mu_p]},
\qquad
\delta\psi_{\mu_1\ldots\mu_{p-1}}=m\lambda_{\mu_1\ldots\mu_{p-1}}
\end{eqnarray}
where $F_{\mu_1\ldots\mu_{p+1}}$ is the strength (\ref{strength}) of
antisymmetric field $\varphi_{\mu_1\ldots\mu_p}$ and
$f_{\mu_1\ldots\mu_p}$ is the strength of field $\psi_{\mu_1\ldots\mu_{p-1}}$.
One can proceed further and fix gauge completely removing field
$\psi_{\mu_1\ldots\mu_{p-1}}$.
Finally we obtain Lagrangian
\begin{eqnarray}
\mathcal{L}&=&
-\frac{1}{(p+1)!}\;
F_{\mu_1\ldots\mu_{p+1}}F^{\mu_1\ldots\mu_{p+1}}
-\frac{m^2}{p!}\varphi_{\mu_1\ldots\mu_p}\varphi^{\mu_1\ldots\mu_p}
\label{Lmf}
\end{eqnarray}
which has no symmetry transformations.
Thus we simplified Lagrangian (\ref{Lm}) and obtained Lagrangian
(\ref{Lmf}) which is commonly used for antisymmetric massive
bosonic field.

\section{Summary}
We have shown that the BRST approach, which was developed earlier
for higher spin field models in flat and AdS spaces, perfectly works
for massless and massive bosonic antisymmetric fields in arbitrary
curved space-time. The obtained theories are reducible gauge models,
the corresponding Lagrangians and gauge transformations are given by
(\ref{L0}), (\ref{GT0}) and (\ref{Lm}), (\ref{GTm}) respectively for
massless and massive theories. In both the theories the order of
reducibility grows with the value of the rank of the antisymmetric
field\footnote{In massless case this statement obviously corresponds
to structure of gauge transformations in conventional formulation.
However, in massive case, the approach under consideration also
leads to reducible gauge models.}. In the massive case we
automatically get a formulation with appropriate St\"uckelberg
fields. Like all the Lagrangians constructed on the base of the BRST
approach, the obtained Lagrangians possess more rich gauge symmetry
and contain more fields in comparison with those which are commonly
used for description of the antisymmetric fields. Due to the
presence of the additional symmetry it is possible to get various
intermediate Lagrangian formulations (like (\ref{Lm-1})) for massive
and massless antisymmetric field theories by partial gauge fixing
and eliminating some of the auxiliary fields.

\section*{Acknowledgements}
The work of I.L.B, V.A.K and  L.L.R. was partially supported by RF
Presidential grant for LSS, project No.\ 4489.2006.2. The work of
I.L.B and V.A.K was partially supported by the INTAS grant, project
INTAS-05-7928 and the RFBR grant, project No.\ 06-02-16346. The work
of I.L.B was also supported in part by the DFG grant, project No.\
436 RUS 113/669/0-3 and joint RFBR-DFG grant, project No.\
06-02-04012.

\end{document}